\documentclass[intlimits,twoside,a4paper]{article}
\usepackage[eqsecnum]{cmpj3}
\usepackage[T1]{fontenc}
\usepackage[utf8]{inputenc}
\usepackage{bm}
\articletype{Regular article}
\Keywords{Critical phenomena;
Equilibrium lattice models; Nonequilibrium lattice models
Finite-size scaling; Monte Carlo methods}

\usepackage{graphicx}
\usepackage[greek,ukrainian,english]{babel}
\usepackage{teubner}
\usepackage{psfrag}

\DeclareMathSymbol{\nsum}{\mathop}{largesymbols}{"58}
\DeclareMathSymbol{\nprod}{\mathop}{largesymbols}{"59}
\DeclareMathSymbol{\nint}{\mathop}{largesymbols}{"5A}

\begin{document}
\title{Finite-Size Scaling of the majority-voter model above the upper critical dimension}
\author[C. Chatelain]{C. Chatelain\orcid{0000-0002-5374-1799}}
\address{Universit\'e de Lorraine, CNRS, LPCT, F-54000 Nancy, France}
\date{\today}

\maketitle 
\begin{abstract}
The majority-voter model is studied by Monte Carlo simulations on hypercubic lattices
of dimension $d=2$ to 7 with periodic boundary conditions. The critical exponents
associated to the Finite-Size Scaling of the magnetic susceptibility are shown to be
compatible with those of the Ising model. At dimension $d=4$, the numerical data are
compatible with the presence of multiplicative logarithmic corrections. For $d\ge 5$,
the estimates of the exponents are close to the prediction
$d/2$ when taking into account the dangerous irrelevant variable at the
Gaussian fixed point. Moreover, the universal values of the Binder cumulant are also
compatible with those of the Ising model. This indicates that the upper critical
dimension of the majority-voter model is not $d_c=6$ as claimed in the
literature, but $d_c=4$ like the equilibrium Ising model.
\end{abstract}

\section{Introduction}
The Renormalization Group Theory, pioneered by Wilson and Fisher
among others, has provided a deep understanding of the critical
behavior of statistical models, such as the Ising model, below their
upper critical dimension $d_c$~\cite{Wilson}. Above $d_c$, the
situation is much simpler and the critical exponents take the values
predicted by mean-field theory. However, the correct Finite-Size Scaling
of thermodynamic averages above the upper critical dimension has been
clarified only recently~\cite{Berche22}.
\\

The critical behavior of the $d$-dimensional Ising model is described
in the continuum limit by the Landau-Ginzburg action~\cite{Binney,Amit,Tauber}
	\begin{equation}
	S[\phi]=\nint\big[|\nabla\phi|^2+r\phi^2+u\phi^4]d^dx.
	\label{Action}
	\end{equation}
Since the action is dimensionless, one can determine the scaling dimensions
of the field and of the couplings by power-counting. The scaling
dimension of $\phi$ is $x_\phi=(d-2)/2$ and therefore, the dimension of
the coupling $u$ is $y_u=d-4x_\phi=4-d$. It follows that the quartic
term is relevant for $d<4$, marginal at $d=4$ and irrelevant at $d>4$.
Below the upper critical dimension $d_c=4$, the critical exponents
can only be estimated using the full machinery of the Renormalization Group (RG).
In contrast, above the upper critical dimension $d_c=4$, one expects
the critical behavior to be governed by the Gaussian fixed point
corresponding to $u=0$. It turns out that the coupling $u$ is a
dangerously irrelevant variable and should be taken into account~\cite{Fisher}.
From the RG flow equations, it can be shown that under a rescaling $x
\rightarrow x/b$ the singular part of the free energy density behaves as
	\begin{equation}
	f(r,u)=b^{-d}f\big(b^2(r+\alpha u)
	-b^{4-d}\alpha u,b^{4-d}u\big).
	\end{equation}
It follows that the critical exponents are not those of the Gaussian fixed
point but the mean-field exponents.
The standard hyperscaling relation holds only at $d=d_c$ and is violated above.
Finite-Size Scaling is also affected by the dangerous irrelevant coupling
$u$. With periodic boundary conditions, the quartic term of the action
Eq.~\ref{Action} involves the volume $V$ of the system in Fourier space:
	\begin{equation}
	S=\nsum_{\vec k} (k^2+r)|\phi_{\vec k}|^2
	+{u\over V^2}\nsum_{\vec k_1,\vec k_2,\vec k_3}
	\phi_{\vec k_1}\phi_{\vec k_2}\phi_{\vec k_3}
	\phi_{-\vec k_1-\vec k_2-\vec k_3}.
	\end{equation}
At the pseudo-critical point $r=0$, the Finite-Size Scaling of the magnetic
susceptibility is dominated by the contribution of the $k=0$
mode~\cite{Brezin}:
	\begin{equation}
	\langle\phi^2\rangle\sim{\nint \phi_0^2e^{-{u\over V}\phi_0^4}d\phi_0
	\over \nint e^{-{u\over V}\phi_0^4}d\phi_0}
	\sim V^{1/2}=L^{d/2}.
	\end{equation}
It is therefore anomalous with a divergence $L^{d/2}$ with the lattice size $L$
and not $L^{\gamma/\nu}$ with $\gamma/\nu=2$ as expected at the Gaussian Fixed
Point. Similarly, the correlation length scales with the lattice size as
$\xi\sim L^{d/d_c}$~\cite{Brezin,Jones}. A coherent Finite-Size Scaling theory
has been presented based on the new exponent \textgreek{\coppa}$=d/d_c$
for $d>d_c$~\cite{Kenna}.
The later predicts that the magnetic susceptibility diverges with the lattice
size with an exponent \textgreek{\coppa}$\gamma/\nu=d/2$.
This theory has recently been extended to quantum phase transitions~\cite{Kai}.
Despite some indications that the same Finite-Size Scaling holds with free
boundary conditions, the problem is not completely settled~\cite{Berche2,Berche22}.
\\

The above discussion concerns only the Ising model but it is believed
to be more general. In particular, the Finite-Size Scaling of percolation
above its upper critical dimension $d_c=6$ has been analyzed in the same
way~\cite{Kenna2}. In this paper, we are interested in the majority
voter model which has the peculiarity, as percolation, of not being
described by a Hamiltonian. This model is nevertheless believed to
belong to the universality class of the Ising model. However,
it has been claimed that the upper critical dimension is not $d_c=4$
but $d_c=6$~\cite{Yang}. We performed extensive Monte Carlo simulations
and compared the Finite-Size Scaling of the magnetic
susceptibility of the majority-voter and Ising models.
In the first section, the two models are more precisely defined
and details on the Monte Carlo simulations are given. The numerical
results are analyzed in the second section. Conclusions follow.

\section{Models and simulation details}
In this study, we consider hypercubic lattices of dimension $d$
ranging from $d=2$ to 7.
Each node $i$ of this lattice is occupied by a classical Ising spin
that can take two possible values $\sigma_i=\pm 1$. The Ising
model is defined by the Hamiltonian
	\begin{equation}
	H(\{\sigma\})=-\nsum_{(i,j)} \sigma_i\sigma_j
	\end{equation}
where the sum extends over all pairs of neighboring sites of the lattice.
At equilibrium with a thermal bath at the temperature $T=1/k_B\beta$,
the probability of a given spin configuration $\{\sigma_i\}_i$ is given
by the celebrated Boltzmann distribution
	\begin{equation}
	\wp_{\rm Eq.}(\{\sigma\})={1\over{\cal Z}}e^{-\beta H(\{\sigma\})}.
	\end{equation}
One can find different Markovian dynamics whose stationary distribution
is the Boltzmann distribution $\wp_{\rm Eq.}$. It is convenient to
impose the detailed balance condition
	\begin{equation}
	\wp_{\rm Eq.}(\{\sigma\})W(\{\sigma\}\rightarrow \{\sigma'\})
	=\wp_{\rm Eq.}(\{\sigma'\})W(\{\sigma'\}\rightarrow \{\sigma\})
	\label{DetailedBalance}
	\end{equation}
which is satisfied by the Glauber transition rates consisting
in single spin flips
	\begin{equation}
	W(\{\sigma\}\rightarrow \{\sigma'\})
	={1\over N}\nsum_i \omega(\sigma_i)\ \!\delta_{\sigma_i',-\sigma_i}
	\nprod_{j\ne i}\delta_{\sigma_j',\sigma_j}
	\label{LocalDyn}
	\end{equation}
with~\cite{Glauber}
	\begin{equation}
	\omega(\sigma_i)={1\over 2}\big[1-\sigma_i\tanh\big(\beta\nsum_j\sigma_j\big)\big]
	\label{Glauber}
	\end{equation}
where the sum extends over all neighbors $j$ of site $i$. The dynamics that is generated
by these transition rates can be seen as a local equilibration of the spin
$\sigma_i$ in the effective magnetic field created by its neighbors.
Note that there are several solutions to the detailed balance condition
Eq.~\ref{DetailedBalance}. The Metropolis transition rate is another
solution that is commonly used in Monte Carlo simulations~\cite{Metropolis}.
These two dynamics are slow: the dynamical exponent is $z=2$ away from criticality
and slightly larger than 2 at the critical point. To compute equilibrium properties
by means of Monte Carlo simulations, it is therefore much more efficient
to use cluster algorithms based on non-local spin updates~\cite{Wolff,
Swendsen}.
\\

The majority-voter model is another Markovian dynamics for Ising spins on a lattice.
As the Glauber dynamics, it consists in single-spin flips, i.e. the transition
rates are of the form Eq.~(\ref{LocalDyn}) but with
	\begin{equation}
	\omega(\sigma_i)={1\over 2}\big[1-\lambda\sigma_i S\big(\nsum_j\sigma_j\big)\big]
	\label{Majority}
	\end{equation}
where $S(x)$ is the sign function defined by $S(x)=1$ for $x>0$, $-1$ for $x<0$,
and 0 for $x=0$. Note that in Ref.~\cite{Yang}, the parameter $\lambda$ is
denoted $\tanh\beta_T$. In contrast to Glauber dynamics, the majority-voter model
does not satisfy the detailed balance condition Eq.~\ref{DetailedBalance}.
No Hamiltonian can be associated to this model and the parameter $\lambda$
is not related to any temperature.
The majority-voter dynamics can be used to study the spreading of opinion
in a population~\cite{Redner}. Each node $i$ of the lattice is associated
to a voter and the spin $\sigma_i$ to his answer to a binary question.
At each time, each voter adapts his choice according to the majority opinion
of his neighbors. Note that the majority-voter model should not be confused
with the voter model~\footnote{In the voter model, each voter chooses one of its
neighbours and adopts his choice.}. The Ising-Glauber and majority-voter models
are special cases of the more general transition rates
	\begin{equation}
	\omega(\sigma_i)={1\over 2}\big[1-\lambda\sigma_i
	\tanh\big(\beta\nsum_j\sigma_j\big)\big].
	\label{General}
	\end{equation}
The Ising-Glauber model is recovered with the choice $\lambda=1$ and
the majority-voter model with $\beta\rightarrow +\infty$. As a consequence,
the majority-voter model appears as an Ising-Glauber model at zero temperature
with an additional noise.

Even though the majority-voter model cannot be associated with any Hamiltonian,
it is believed that the averages computed in the stationary distribution display
the same critical behavior as the equilibrium Ising model~\cite{Grinstein}.
Measurements of the static critical exponents of the 2D majority-voter model
by Monte Carlo simulations indeed support this idea~\cite{Oliveira, Sastre2}.
The dynamical exponent $z$ and the initial critical slip exponent $\theta$ were
also shown to be compatible with those of the Ising model~\cite{Mendes}.
A more recent Monte Carlo simulation of the 3D majority-voter model also found
critical exponents in the Ising universality class~\cite{Sastre,Nascimento}.
However, an extensive Monte Carlo simulations in dimensions $d=2$ to 7
reached the conclusion that the upper critical dimension of the
majority-voter model is not $d_c=4$, like the Ising model, but
$d_c=6$~\cite{Yang}.
\\

In this work, the critical behavior of the Ising-Glauber and majority-voter
models is studied by Monte Carlo simulations. Much more accurate estimates
of the critical exponents of the Ising model could have been computed with
cluster algorithms. However, our goal is here to study both models
with local dynamics and with the same number of Monte Carlo iterations
in order to compare exponents with similar error bars. The code
was parallelized with the Cuda language and run on GPUs Nvidia Tesla P100
and GTX 1080. Hypercubic lattices of dimension $d=2$ to 7 were considered
with periodic boundary conditions.
The largest lattice sizes that could be reached are
$1024$ in 2D, 120 in 3D, 36 in 4D, 18 in 5D, 12 in 6D and 8 in 7D.
$10^5$ iterations were performed to thermalize the system and $10^6$
iterations were used to compute the averages. Several independent
simulations were performed and the error was estimated as
$\sigma/\sqrt{N}$ where $\sigma$ is the standard deviation among
the $N$ independent simulations.

\section{Numerical results}
On Fig.~\ref{fig1}, the magnetic susceptibility
	\begin{equation}
	\chi=L^d\big[\langle m^2\rangle-\langle m^2\rangle\big]
	\end{equation}
is plotted versus the parameter $\lambda$ of the majority-voter model
and $\beta$ of the Ising model in dimension $d=5$. For each lattice size,
the pseudo-critical parameters $\lambda_c(L)$ and $\beta_c(L)$ were estimated
as the location of the maximum of the susceptibility.
To improve the accuracy, a quadratic fit of the data is first performed
over the points for which $\chi\ge 0.7\max\chi$ and the maximum is computed
from the parameters of the fit. On Fig.~\ref{fig2}, the magnetic susceptibility
at the pseudo-critical point is plotted versus the lattice size. The critical
exponent \textgreek{\coppa}$\gamma/\nu$ is estimated from a simple power-law
fit over all data. The estimates are collected in Table~\ref{Tab1}. The critical
exponents \textgreek{\coppa}$\gamma/\nu$ of the majority-voter and Ising models
are compatible within error bars, except at $d=4$ where the two error bars do
not overlap but are very close to each other (the distance between them is $0.03$).
Note that $d=4$ is the upper critical dimension of the Ising model
and, as will be shown in this paper, of the majority-voter model.
RG predicts the presence of logarithmic corrections in the scaling behavior
of the $\phi^4$ model with temperature~\cite{Binney,Amit,Tauber}. These corrections
were shown to be present in Finite-Size Scaling too~\cite{Kenna3}.
In dimension $d>4$, the critical exponents
\textgreek{\coppa}$\gamma/\nu$ of both the majority-voter and Ising model
are close to the prediction $d/2$ at the Gaussian fixed point when
taking into account the dangerous irrelevant variable, although not
compatible within error bars.

\begin{figure}
    \centering
    \psfrag{beta}{$\beta$}
    \psfrag{lambda}{$\lambda$}
    \psfrag{Chi}{$\chi$}
    \caption{Susceptibility $\chi$ of the 5D Majority-Voter model (left)
    and of the 5D Ising model (right) versus the parameters $\lambda$ and
    $\beta$. The different curves correspond to different lattice sizes.}
    \label{fig1}
\end{figure}

\begin{figure}
    \centering
    \psfrag{L}{$L$}
    \psfrag{Chi}{$\chi$}
    \includegraphics[width=0.65\textwidth]{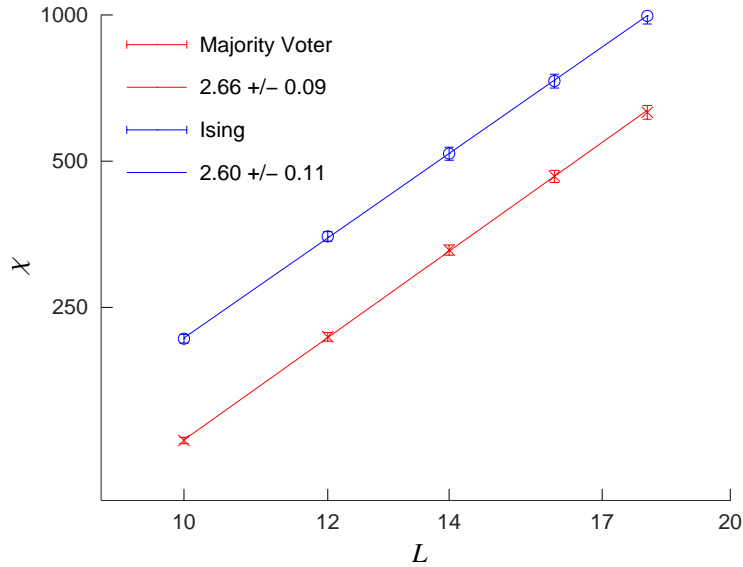}
    \caption{Maximum of the susceptibility of the 5D Majority-Voter model (cross)
    and of the 5D Ising model (circle) versus the lattice size $L$.
    The solid lines are power-law fits of the data. The estimated critical
    exponents $\gamma/\nu$ are indicated in the legend.
    }
    \label{fig2}
\end{figure}

\begin{figure}
    \centering
    \psfrag{d}{$d$}
    \psfrag{phi4}{$\phi^4$ theory}
    \psfrag{g/v}{\textgreek{\coppa}$\gamma/\nu$}
    \includegraphics[width=0.65\textwidth]{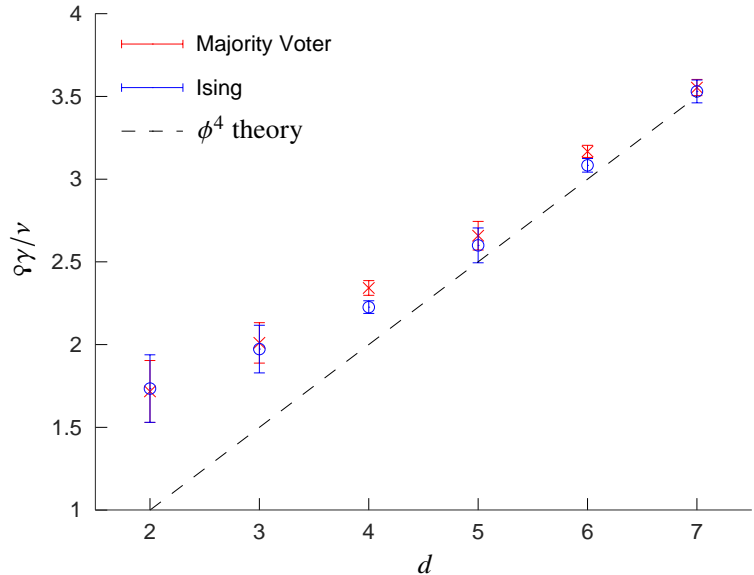}
    \caption{Critical exponent \textgreek{\coppa}$\gamma/\nu$ obtained from
    the Finite-Size Scaling of the susceptibility of the Majority-Voter model
    and the Ising model versus the dimension $d$ of the lattice.
    The dashed line is the prediction $d/2$ at the Gaussian fixed point
    with dangerous irrelevant variables that is expected to hold
    above the upper critical dimension.}
    \label{fig3}
\end{figure}

\begin{table}[htb]
\caption{
Estimates of the critical exponent \textgreek{\coppa}$\gamma/\nu$ for the
Majority Voter model (left) and the Ising model (center) on an hypercubic
lattice of dimension $d$. Known values for the Ising model are given in
the right column. At $d=3$, the value is a recent estimate by Functional
RG~\cite{Balog}. Other values are exact. At $d=d_c=4$, multiplicative
logarithmic corrections are present.}
\begin{center}
\begin{tabular}{l|lll}
$d$ & Majority Voter & Ising & Known\\
\hline
2	& 	 1.72(19) & 1.73(20) & 1.75 \\
3   &    2.01(12) &  1.97(14) & 1.96370(20)\\
4   &    2.34(4) & 2.23(4) & 2 (+log)\\
5   &    2.66(9) & 2.60(11) & 2.5\\
6   &    3.17(4) &  3.08(4) & 3 \\
7   &    3.55(5) & 3.53(7) & 3.5\\
\end{tabular}
\end{center}
\label{Tab1}
\end{table}

\begin{figure}
    \centering
    \psfrag{log(L)}{$\ln L$}
    \psfrag{Chi/L2}{$\chi/L^2$}
    \includegraphics[width=0.65\textwidth]{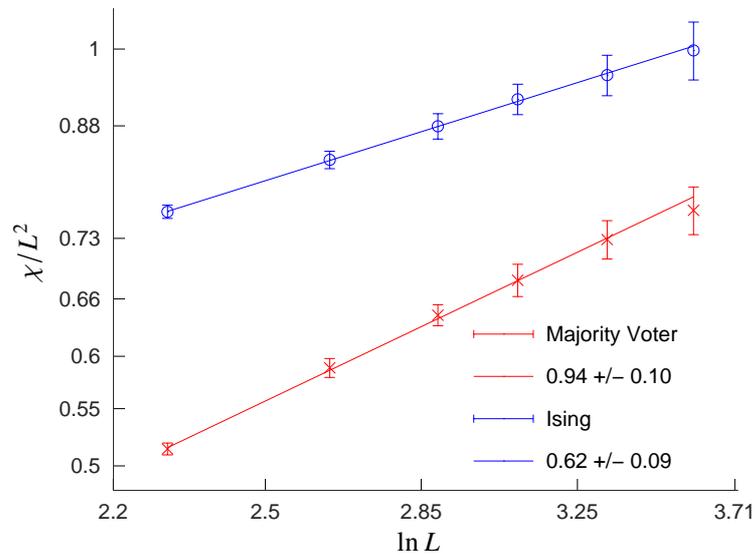}
    \caption{Maximum of the susceptibility of the 4D Majority-Voter model (cross)
    and of the 4D Ising model (circle) divided by $L^2$ versus the
    logarithm $\ln L$ of the lattice size $L$.
    The solid lines are power-law fits of the data. The estimated
    exponents $\hat\gamma/\hat\nu$ are indicated in the legend.
    }\label{fig5}
\end{figure}

As already discussed above, the critical behavior of the magnetic susceptibility
of the Ising model is expected to involve multiplicative logarithmic corrections
at the upper critical dimension $d_c=4$. It can be inferred that the Finite-Size
Scaling of the magnetic susceptibility at its maximum is of the form
	\begin{equation}
	\chi\sim L^{d/2}(\ln L)^{\hat\gamma/\hat\nu}
	\end{equation}
where $\hat\gamma/\hat\nu$ is known to be $1/2$ for the Ising model
at the critical temperature $\beta_c$~\cite{Kenna3}.
To test this scaling for the majority-voter model, the quantity $\chi/L^2$
has been plotted versus $\ln L$ with a logarithmic scale on Fig.~\ref{fig5}.
A very nice power-law behavior can be observed for both
the majority-voter model and the Ising model. The exponent $\hat\gamma/\hat\nu$
is estimated to be $0.94(10)$ for the majority-voter model and $0.62(9)$ for the
Ising model. For the Ising model, the known value $\hat\gamma/\hat\nu=1/2$
is slightly outside of the error bar of our estimate. Note also that, for
both the majority-voter and the Ising models, the magnetic susceptibility
can also be fitted with logarithmic corrections in dimensions $d>4$
but with an exponent $\hat\gamma/\hat\nu$ decreasing with $d$ ($0.41(22)$
at $d=5$ and $0.30(6)$ at $d=6$ for the majority-voter model).

\begin{figure}
    \centering
    \psfrag{beta}{$\beta$}
    \psfrag{lambda}{$\lambda$}
    \psfrag{U}{$U$}
    \includegraphics[width=0.65\textwidth]{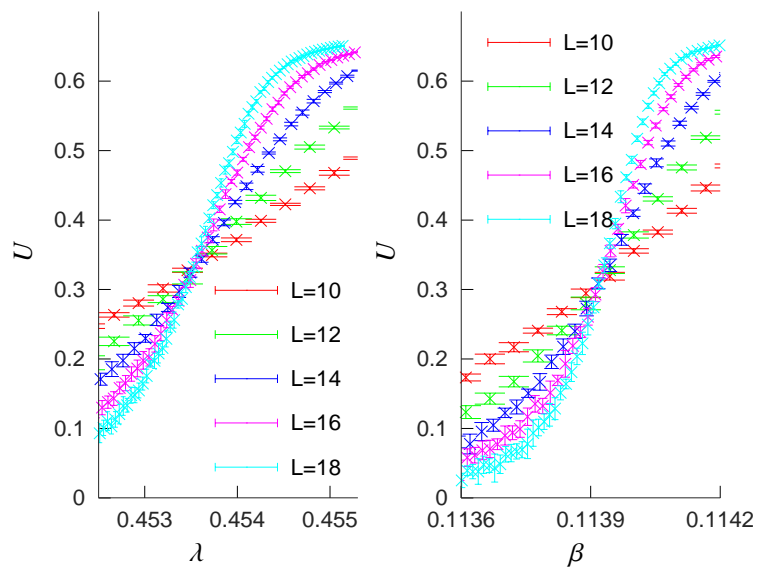}
    \caption{Binder cumulant $U$ of the 5D Majority-Voter model (left)
    and of the 5D Ising model (right) versus the parameters $\lambda$ and
    $\beta$. The different curves correspond to different lattice sizes.}
	\label{fig4}
\end{figure}

On figure~\ref{fig4}, the Binder cumulant
	\begin{equation}
	U=1-{\langle m^4\rangle\over 3\langle m^2\rangle^2}
	\end{equation}
is plotted versus the parameters $\lambda$ and $\beta$ in dimension $d=5$.
The curves for different lattice sizes are expected to cross at the critical
parameters $\lambda_c$ and $\beta_c$ in the thermodynamic limit
and the value of the Binder cumulant at the crossing points is expected
to be universal. For all dimensions, very similar values are observed
for the majority-voter and Ising models. In dimensions $d=5$,
we have estimated the universal values to be $U_*(\infty)\simeq 0.27(7)$
for the majority-voter model and $U_*(\infty)\simeq 0.30(6)$ for the Ising
model. Estimates at dimensions $d=2$ to 7 are given in Table~\ref{Tab2}.
They were obtained by a linear interpolation of the Binder cumulant
to estimate more accurately the crossing points. The value $U_*(L)$ at
the crossing is then fitted with the law $U(L)=U_*(\infty)+b/L$.
The absence of error bars for the largest lattice sizes is due to a
too small number of points in the fit.
The accuracy on the Binder cumulant is unfortunately
not sufficient to estimate $\nu$ from the Finite-Size Scaling
of ${d\over d\lambda}U$ and ${d\over d\beta}U$.

\begin{table}
\caption{
Estimates of the universal Binder cumulant $U_*(\infty)$ for the Majority Voter model
(left) and the Ising model (right) on an hypercubic lattice of dimension $d$.}
\begin{center}
\begin{tabular}{l|ll}
$d$ & Majority Voter & Ising \\
\hline
2	& 	 0.62(4) & 0.62(4) \\
3   &    0.47(6) &  0.46(6) \\
4   &    0.36(2) & 0.33(2) \\
5   &    0.27(7) & 0.30(6) \\
6   &    0.31(5) &  0.29(3) \\
7	& 	 0.40 	 & 0.33 \\
\end{tabular}
\end{center}
\label{Tab2}
\end{table}

\section*{Conclusions}
In conclusion, we have provided numerical evidences that the majority-voter
model belongs to he universality class of the equilibrium Ising model for
any dimension $2\le d\le 7$. The immediate consequence is that the upper
critical dimension of the majority-voter model is $d_c=4$, like the Ising
model. Above the upper critical dimension, the Finite-Size Scaling of the
magnetic susceptibility is indeed close to the prediction $\chi\sim L^2$
at the Gaussian fixed point with a dangerous irrelevant variable.
For both the majority-voter and Ising models,
the small deviation to this law may be attributed to
the too small lattice sizes that could be reached, and therefore to
scaling corrections. At the upper critical dimension, multiplicative
logarithmic corrections are present in both the majority-voter and Ising
models but with a different exponent. Note that this is not a proof that
$d_c=4$ since the data can also be fitted with logarithmic corrections
for $d>4$, although with smaller exponents.

\section*{Acknowledgments}
The author would like to warmly congratulate Bertrand Berche
for his 60th birthday. This paper was submitted to the Festschrift
in his honor. The author is also grateful to Kai Phillip Schmidt for discussions.
The numerical simulations of this work were performed at the meso-center
eXplor of the universit\'e de Lorraine under the project 2018M4XXX0118.

\section{Bibliography}
\bibliographystyle{cmpj}

\end{document}